\documentclass[12pt]{article}
\def\bib{\bibitem}
\def\ql{\lq\lq}
\def\qr{\rq\rq}
\def\ft{\footnote}
\def\sp{$\;$}
\def\bq{\begin{quote}}
\def\eq{\end{quote}}
\begin{document}
\begin{flushright}
\today
\end{flushright}
\begin{center}
\Large{\bf The Reasonable Effectiveness of Mathematics in the Natural Sciences}\\
\vspace{.5cm}
\large{Alex Harvey}${}^{a)}$ \\
\vspace{.15cm}
\normalsize
Visiting Scholar \\
New York University \\
New York, NY 10003 \\
\vspace{.8cm}
\end{center}
\begin{abstract}
Mathematics and its relation to the physical universe have been the topic of speculation since the days of Pythagoras.  Several different views of the nature of mathematics have been considered: {\it Realism\/} - mathematics exists and is discovered; {\it Logicism\/} - all mathematics may be deduced through pure logic; {\it Formalism\/}: mathematics is just the manipulation of formulas and rules invented for the purpose; {\it Intuitionism\/}: mathematics comprises mental constructs governed by self evident rules. The debate among the several schools has major importance in understanding what Eugene Wigner called, \ql The Unreasonable Effectiveness of Mathematics in the Natural Sciences.\qr$\,$  In return, this `Unreasonable Effectiveness' suggests a possible resolution of the debate in favor of {\it Realism\/}.  The crucial element is the extraordinary predictive capacity of mathematical structures descriptive of physical theories.\ft{It is a pleasure to dedicate this paper to Josh Goldberg, long a valued member of the community of Generl Relativists and a good friend.}
\end{abstract}
\vspace{.5cm}
\begin{flushleft}
\end{flushleft}
\vspace{.8cm}
\section{Introduction}
In an essay which appeared in his collection, {\it Symmetries and
Reflections\/}, Wigner \cite{wigner} explored the connection
between mathematics and science.  He wondered that, 
\bq
\ql $\dots$the enormous
usefulness of mathematics in the natural sciences is something
bordering on the mysterious and that there is no rational
explanation for this.\qr
\eq
And, later,
\bq
\ql $\ldots$ the mathematical formulation of the physicist's often crude experience leads in an uncanny number of cases to an amazingly accurate description of a large class of phenomena.\qr
\eq
A similar view was expressed by Bochner\ft{It is noteworthy that Wigner was a Nobel Laureate in physics and Bochner, a distinguished mathematician.  The talent for the two skills does not seem to be the same.  Only a few gifted individuals such as Archimedes, Newton, or Gauss had supreme talents in both.  This difference in talents can be observed in college upperclassmen.  Expert knowledge and superior skill in the {\em use\/} of mathematics is the hallmark of the gifted physicist.} \cite{bochner}
\bq
\ql What makes mathematics so effective when it enters science is a mystery of mysteries and the present book wants to achieve no more than explicate how deep this mystery is.\qr
\eq
And again, in Chapter 22 of the {\it London Philosophical Study Guide\/} \cite{lpsg}.
\bq\ql
Mathematical reality is in itself mysterious: how can it be highly abstract and yet applicable to the physical world? How can mathematical theorems be necessary truths about an unchanging realm of abstract entities and at the same time so useful in dealing with the contingent, variable and inexact happenings evident to the senses?\qr\eq

None mention the central mystery, which is the ability of an {\it appropriate \/} formulation to predict phenomena yet undiscovered.  Indeed, as often noted, the principal problem in progress in physics is to find this formulation.  It is our purpose to explore this \ql mysterious\qr$\;$ relation between physical theories and their mathematical formulation and suggest an answer.  Firstly an understanding of mathematics is essential.
\section{Mathematics}
Mathematics arose from that most fundamental of mathematical activities - counting objects.  In the most ancient of times, a person might {\em enumerate\/} objects of the same kind - say goats or sheaves of wheat and if he wanted to record the results, would have to invent a system of symbols to identify and differentiate two quantites.  We can be sure that he would use the same symbol system to record different quantities of apples.  He would quickly learn to add two collections of objects.  This could well have happened anywhere groups of people progressed beyond being hunter-gatherers. 

This simple skill evolved to the sophisticated arithmetic of the Babylonians \cite{neugebauer}.  They exhibited great ingenuity in their ability to use tables of squares to solve various arithmetic problems and could solve quadratic and cubic equations.  Similarly, a direct descent for spatial relations can be discerned. The Pharaonic Egyptians devised an elementary geometry for purposes of resurveying fields after the recession of the annual Nile floodwaters.  This skill was elaborated to the extent that construction of temples and pyramids became possible.  To accomplish all this, a unit of length was defined.  In this way geometry was conflated with arithmetic.  Despite their achievements, neither the Babylonians nor the Egyptians considered their mathematics to be more than a computational scheme.  Their mathematics was devoted primarily to mensuration, computing taxes, keeping accounts and making astronomical calculations.  They had no concepts of theorems or proofs.  This was left to the Greeks. 

The genius of the Greeks, beginning with the Pythagorean school was to discern an abstract coherent structure beneath the arithmetic of the Babylonians and Egyptians.  They introduced the concept of {\em theorem and proof\/}.  In this way, from the simple act of counting and the drawing of geometric figures, an elaborate system of mathematics was constructed in Classical Greece.  

The Pythagoreans \cite{allen1} left no written record of their accomplishments.  These are known largely through the writings of later Greek philosophers \cite{klein}.  They were besotted with number {\it per se\/} effectively founding number theory.  But, it was their belief in the esoteric character of mathematics and its relation to the physical universe which is of interest here.  They believed that all that could be known of the physical universe would be known through number.  Of the Pythagoreans, Aristotle  commented \cite{ross}, 
\bq
\ql Contemporaneously with these philosophers and before them, the so-called Pythagoreans, who were the first to take up mathematics, not only advanced this study, but also having been brought up in it they thought its principles were the principles of all things. Since of these principles numbers are by nature the first, and in numbers they seemed to see many resemblances to the things that be modeled on numbers, and numbers seemed to be the first things in the whole of nature, they supposed the elements of numbers to be the elements of all things, and they exist and come into being - more than in fire and earth and water $\ldots$
 all other things seemed in their whole nature to hold heaven to be a musical scale and a number.\qr
\eq

Many centuries later, Galileo \cite{galileo} put it more precisely (and famously). 
\bq
\ql Philosophy is written in this grand book--I mean the universe--which stands continually open to our gaze, but it cannot be understood unless one first learns to comprehend the language and interpret the characters in which it is written. It is written in the language of mathematics, and its characters are triangles, circles, and other geometrical figures, without which it is humanly impossible to understand a single word of it; without these, one is wandering around in a dark labyrinth.\qr
\eq
Whether his view is Pythagorean, that is, does he imply an identity between the mode of description and the object described or that the mode and object are distinct, is not clear.

A similar sentiment, with clearer distinction, was expressed some three centuries later by Weyl \cite{weyl} in connection with the \ql language\qr\sp appropriate for application to gravitation theory.
\bq\ql As everyone has to work hard learning language and writing before he can use them freely for expressing his thoughts, so here too the only way to shift the load of formulas from one's own shoulders is to be well acquainted with tensor analysis so that one can turn unimpeded by formalities to the true problems:  Insight into the nature of space, time and matter insofar as they contribute to the construction of the objective reality.\qr\eq
\section{Nature of Mathematics}
For the scientist, mathematics is a tool; for the mathematician, it is an end in itself; for the philosopher, the philosophy of mathematics is, as K\"{o}rner \cite{korner1} puts it, 
\bq
\ql ... not mathematics.  It is a reflection on mathematics.\qr
\eq  
The {\it nature\/} of mathematics has been debated by philosophers\ft{For a concise introduction and useful bibliography, see the Wikipedia article, \ql The Philosophy of Mathematics\qr.  For a detailed introduction and discussion, consult K\"{o}rner's \cite{korner2}\ql {\it The Philosophy of Mathematics\/}\qr.}  
beginning with the Pythagoreans.  They considered numbers to be inextricably connected with the physical world.  Plato refined Pythagorean ideas to a scheme of {\it ideal} abstract structures mirrored in the immediate world as {\em discovered\/} through human activity.  Subsequent modifications led to {\it Realism\/} in which the severe concept of ideals is modified.  Simply put, mathematical structures have an independent existence and are uncovered by observation and ratiocination.  This view has been held by many mathematicians such as Paul Erd\"{o}s who often referred to an imaginary \ql Book\qr$\,$ in which God had written down all the \ql beautiful\qr$\,$  mathematical proofs.  
This was expressed more picturesquely by Everett \cite{everett}
\bq 
\ql In the pure mathematics we contemplate absolute truths which existed in the divine mind before the morning stars sang together, and which will continue to exist there when the last of their host shall have fallen from heaven.\qr
\eq
A more temperate exposition of the same view was expressed by the distinguished British mathematician, Hardy \cite{hardy}, 
\bq 
\ql I have myself always thought of a mathematician in the first instance as an observer, who gazes at a distant range of mountains and writes down his observations.\qr
\eq

In the late 19th Century, a radically new approach by Frege \cite{frege}, subsequently termed {\it Logicism\/} claimed that all mathematics was just a logical structure.
\bq
\ql Arithmetic thus becomes simply a development of logic, and every proposition of arithmetic a  law of logic albeit a derivative one.  To apply arithmetic in the physical sciences is to bring logic to bear on observed facts; calculation becomes deduction.  The laws of number will not, $\ldots$, need to stand up to practical tests if they are to be applicable to the external world; for in the external world, in the whole of space and all that therein is, there are no concepts, no properties of concepts, no numbers.  The laws of number, therefore, are not really applicable to external things; they
are not laws of nature.\qr
\eq
A modified view was earlier given by Kronecker \cite{kronecker}.
\bq
\ql God created the numbers, all the rest is the work of man.\qr
\eq
This avoids the extreme difficulty of establishing the system of numbers by purely logical arguments encountered later by Whitehead and Russell in their {\it Principia Mathematica\/} \cite{whitehead}.  They do not establish that $1+1=2$ until page 362 of Volume 1.

The concept of {\it Formalism\/} is due to Hilbert who described it simply as\ft{Though widely quoted, the attribution of this comment to Hilbert is apocryphal.} \cite{rose}.
\bq
\ql Mathematics is a game played according to certain simple rules with meaningless marks on paper.\qr
\eq
Wigner, early in his essay, in the section titled {\it What is Mathematics?\/}, provides the same view of mathematics.
\bq
\ql$\ldots$ mathematics is the science of skillful operations with concepts and rules invented for just this purpose.\qr
\eq

{\it Intuitionism\/}, created by the Dutch mathematician Brouwer \cite{kleene},  proposes that mathematical objects are mental constructions governed by self-evident laws. 

In a class by itself is the school of thought that considers the brain to have been hard-wired with the number concept at birth\ft{For an example of this approach see Dahaene \cite{dahaene}.}

The various schools of thought have enabled intense debates among philosophers of science and mathematics.  However, a conclusive argument strongly in favor of Realism can be constructed out the interaction of mathematics and physics,  This will be described below.
\section{The Practice of Physics}
The physicist attempts to understand physical universe.  The endeavor is enormously facilitated through employment of Galileo's \ql language of mathematics\qr. The use of mathematics entails representing each physical entity of interest with a mathematical object.  This is done by {\em mapping\/} these entities {\it uniquely\/} onto a {\it mathematical\/} concept.  A velocity is a polar vector; a moment is an axial vector; time is a one-dimensional manifold; flat space-time is a four-dimensional manifold with a Minkowski metric.  The process is a species of conceptual {\it isomorphism}.  It can't be a homeomorphism if ambiguity is to be avoided.  Once this mapping has been completed, the mathematics takes over completely; it has a life of its own.  The rules of the particular mathematical structure govern.  There is no longer any physics {\it per se\/}.  At the end of the calculation, the results are mapped back onto the physical universe.  

At intermediate steps it should be possible to map the mathematics back onto the physics.  That is, it should be possible to interpret every mathematical symbol with a physical construct.  Dirac \cite{dirac1} expressed it this way, 
\bq\ql The new scheme [the representation of quantum states and variables]  becomes a precise physical theory when all the axioms and rules of manipulation governing the mathematical quantities are specified and when in addition certain laws are laid down connecting physical facts with the mathematical formalism, so that from any given physical conditions equations between the mathematical quantities may be inferred and vice versa. In an application of the theory one would be given certain physical information, which one would proceed to express by equations between the mathematical quantities.  One would then deduce new equations with the help of the axioms and rules of manipulation and would conclude by interpreting these new equations as physical conditions.\qr\eq 

The \ql Unreasonable Effectiveness\qr$\,$ is two-fold. Firstly, there is the incredible richness of phenomena which can be accurately described by a mathematical structure.  Maxwell's equations furnish the means to accurately describe electromagnetic phenomena from power transformers to radio transmission.  Einstein's equations of general relativity provide the basis for calculating gravitational phenomena from remarkably accurate solar system ephemerides to the gravitational radiation from pulsar PSR1913+16.  The wealth of accurate description of atomic phenomena by the Schr\"{o}dinger equation needs no elaboration.  This brief list could be extended dramatically. 

Secondly, and of decisive importance, the mathematics may describe phenomena completely unforeseen when the problem was originally formulated. think of  the Dirac equation for the electron.  What Dirac \cite{dirac2} said about it later is precisely to the point.
\bq 
\ql It was found that this equation gave the particle a spin of
half a quantum.  And also gave it a magnetic moment.  It gave just the
properties that one needed for an electron.  That was really an
unexpected bonus for me, {\em completely unexpected}.\qr
\eq
[Emphasis added]  What he did not know at the time he published the equation [1927] was that, quite remarkably, it also described the positron which had yet to be discovered [1932].  

The Schwarzschild singularity in the solution of the Einstein equations for an isolated point mass is another example.  So was the yet to be observed radio waves in Maxwell's equations.  Physics is full of these surprises.  A less well known far more simple example is that of the exact solution of the simple pendulum.  This is an elliptic integral \cite{appel}, the real value of which corresponds to the given initial conditions.  The imaginary period corresponds to a reversed gravitational field and an initial displacement of the bob which is supplementary to that in the original problem.  The latter is an example of the principle that every prediction of a mathematical formulation of a physical has a \ql real\qr\sp significance.
\section{Reflections}
The practice of physics is based on the concept that the physical universe, locally and globally, evolves in accordance with fixed rules.  This is attested to by the consistent ability of independent observers to replicate each others results.  It is this regularity which admits the use of mathematics to describe these results.  

All of mathematics starts from a set of integers used for counting.  The historical record shows that the ability to count evolved in different areas independently.  The Babylonian and Mayan civilizations, among others, developed this skill. Contrarily, some cultures developed counting no further than "one, two, many." This suggests an objective independence of the integers and that they were {\it discovered\/} rather than created or, as suggested, that the number concept is hard-wired in the brain \cite{dahaene}.  Further, evidence provided by astrophysical observations implies that the laws of physics are invariant with respect to location and independent of era. This implies that the mathematical models created by extraterrestrials (if such exist) are identical, {\it modulo\/} symbology, to those here on earth.  In turn, these imply the universal common existence of mathematics.  

It was noted earlier that the relation between mathematics and physical phenomena is a species of isomorphism.  The phenomena is mapped uniquely onto a mathematical structure and then mapped in the reverse direction.   The extent to which the initial choice of mathematical structure is an accurate mapping is determined by how well the evolving physical phenomena remains isomorphic to the mathematics   The relationship between mathematics and phenomena is reciprocal; {\em each is a representation\/} of the other.
It is in this sense that mathematics has an objective reality.  The problem of the physicist is to find that representation.

The identification of mathematics as objectively existent entails an interesting problem.  
Hawking \cite{hawking} has noted that mathematical structures are subject to the incompleteness theorems of G\"{o}del:
\bq\ql
{\em ... we and our models are both part of the universe we are describing.\/} Thus a physical theory is self referencing, ... . One might therefore expect it to be either inconsistent or incomplete. ...  The theories we have so far are both inconsistent and incomplete.\qr
\eq
This is certainly correct but only of theoretical interest.  There has yet to constructed mathematical models which are precisely descriptive of the phenomena we do observe.  Despite an incredible accuracy of ten parts in a billion, the present formulation of quantum electrodynamics is notoriously lacking in rigor \cite{valenti}; the \ql standard model\qr, however accurate its predictions, is substantially more a protocol than a theory; a quantized version of the theory of general relativity has yet to be found.  In brief, there is nothing yet in the mathematical models of observed physical phenomena to which G\"{o}del's theorems might be applied.  Moreover, our theories are certainly incomplete and will always remain so.  With respect to the cosmos this is obvious.  The strictures imposed by the principle of special relativity limits what information we can collect to the interior of the past light-cone.  That is finite in extent, some 13+ billion years at the present time.  The best that a physicist may hope for is a model which is both a) accurate in description b) possessed of predictive capability and c) mathematically rigorous.  While the success of (a) and (b) have been substantial, this is not the case for (c).  The approach to a final, mathematically rigorous \ql theory of everything\qr\sp is at best, asymptotic. Thus, for the finite future, G\"{o}del's theorem is not a problem.

Physicists do not (with rare exception such as Newton's fluxions, the aforementioned Dirac delta function, or Heaviside's operational calculus) create any of the mathematics they utilize.  The elaboration of existing and discovery of new mathematics is mostly the province of mathematicians.  We thus have a very large apparatus of which only part is presently used by physicists.  This suggested the question \cite{neuenschwander}, \ql Does any piece of mathematics exist for which there is {\em no application whatsoever in physics\/}?\qr\sp The question has been addressed several times, most recently in 2001 \cite{torre} \cite{gottlieb}.  The answers were uniformly no.  This, though suggestive, is not conclusive.  

If either Formalism or Constructivism were accurate characterizations of mathematics then every mathematician could construct his own universe.  There could be as many different universes as there were mathematicians engaged in the enterprise.  The consequences of this are not easily described.  The decisive evidence that Realism is the proper characterization is the {\em predictive\/} capability of the mathematical structures in which physical theories are couched. This strongly infers that mathematics and the reality it describes are both part of the objective universe and await discovery.
\section{Acknowledgments}
Thanks: to (the late) Professor Arthur Komar for stimulating conversations; to Professor Hilail  Gildin for useful information on the philosophy of mathematics and comments on the manuscript; and especially to Professor Engelbert Schucking who not only provided the translation of the quotation of Weyl, page 4, above but also read the manuscript critically.  
\vspace{.5cm}
\newline 
${}^{a)}$Professor Emeritus, Queens College, City University of New York, 
ah30@nyu.edu
\newline 
\end{document}